\documentclass[12pt,draft]{iopart}

\usepackage{amssymb}
\usepackage{iopams}
\usepackage{footnote}

\begin{document}

\title[Molecular Brownian motion and fundamental
1/f noise]{Once again on molecular Brownian motion \\
and related fundamental 1/f\, noise\,: \\
a logical analysis of exact equations}

\author{Yu E Kuzovlev}

\address{Donetsk Institute
for Physics and Technology of NASU, 83114 Donetsk, Ukraine}
\ead{kuzovlev@fti.dn.ua}

\begin{abstract}
The paper contains a simple semi-quantitative analysis %
of a structure of solution to the exact Bogolyubov %
functional equation for a particle interacting with %
ideal gas and driven by an external force, %
in comparison with solutions to model kinetic %
equations for the same system. %
It is shown that the exact equation inevitably %
predicts existence of significant 1/f-type %
fluctuations in mobility of the particle, and %
this result directly extends to particles in %
arbitrary fluid.
\end{abstract}


\vspace{2pc} \noindent{\it Keywords}\,: Dynamical foundations of , of
kinetics, molecular Brownian motion, 1/f-noise, Mobility 1/f
fluctuations

\section{Introduction}

Here I return to the previously considered \cite{pro} random
wandering (``Brownian motion'') of a particle interacting with ideal
gas. Investigation of this system, with the aim to %
obtain complete actual quantitative statistical %
characteristics of the Brownian motion, still remains %
an intriguing mathematical problem for the future. %
The principal preliminary analysis of this problem, %
as well as of microscopic Brownian motion in fluids %
in general, was undertaken in %
\cite{pppro,ppro,tmf,ig,hs,hs1} and even much %
earlier in \cite{i1,i2} (with continuation in %
\cite{p1,p3}), and in respect to various non-fluid %
systems in \cite{bk12,bk3,i3,kmg} and \cite{i2}. %
Basing on exact complete sets of evolution equations %
for many-particle probability distribution functions, %
- the Bogolyubov-Born-Green-Kirkwood-Yvon (BBGKY) %
equations \cite{bog}, - it was argued that long-scale %
statistics of the Brownian motion qualitatively %
differs from the Gaussian one, thus rejecting the %
``law of large umbers''. This means that diffusivity %
and mobility of a molecular Brownian particle (BP) %
have no definite values but instead undero low-frequency %
fluctuations with some 1/f-type spectrum. Although %
that are natural thermal fluctuations, they were %
lost by conventional ``Boltzmannian'' approaches to %
fluid kinetics, since these approaches neglect %
many-particle statistical correlations. But the %
1/f fluctuations appear as soon as one takes %
into account all the infinite chain of many-particle %
statistical correlations. Paradoxically, the latter %
originate from unpredictability of dynamical %
cause-and-consequence correlations between %
inter-particle collisions, as it was explained %
first in \cite{bk12,bk3} and \cite{i1} and later %
in \cite{i2,p1,pppro,tmf,ig,hs1}.  Most %
mathematically rigorous results in this field are %
presented in \cite{ppro} and \cite{tmf} while %
most developed approximate solutions to the %
BBGKY equations in \cite{i1,i2,p1,p3}.

Nevertheless, there is a lack of formally %
simple and at the same time irrefutably convincing %
proofs of the foregoing statements. %
Here I suggest such a proof. It covers not only %
thermodynamically equilibrium Brownian motion but
also (and first of all) non-equilibrium one under %
a constant external force applied to the BP.

\section{The equations and the problem}

The BBGKY hierarchy describing BP in ideal gas were
written out in various representations and discussed
in \cite{pro,ppro,tmf}. Therefore,
please, see firstly Sections 2 and 3 in \cite{pro}.
Then now we can start from the evolution equation
for generating functional \cite{ppro,tmf,bog}
of the cumulant correlation functions:

\begin{eqnarray}
\frac {\partial \mathcal{V}}{\partial t}\,+
\,V\cdot\frac
{\partial \mathcal{V}}{\partial {R}}\,
+\,f \cdot\frac
{\partial \mathcal{V}}{\partial {P}}
\,=\,\nonumber\\
=\, \int \psi(x)\,\, (V-v)\cdot \frac {\partial }{\partial \rho}\,
\frac {\delta \mathcal{V}}{\delta \psi(x)}\,
\,+\,\,\nonumber\\
+\,\,\int  \psi(x)\,\,\Phi^{\prime }(\rho)
\cdot \left(\frac
{\partial }{\partial p}-
\frac {\partial }{\partial P}\right)
 \frac {\delta \mathcal{V}}{\delta \psi(x)}
\,\,-\,\nonumber\\
-\,\,  \frac {\partial }{\partial P} \cdot \int \Phi^{\prime }(\rho)
 \frac {\delta \mathcal{V}}{\delta \psi(x)}
\,\,+\,\label{fev}\\
+\,\nu\, \left[ \int G_m({p})
\,E^{\,\prime}(\rho)\,\,\psi(x)\,\right] %
\cdot\left( V + T\,\frac
{\partial }{\partial P}\right
)\,\mathcal{V}\,\,\equiv\,\nonumber\\
\equiv\, \widehat{\mathcal{L}}\,\mathcal{V}
\,\,\,, \nonumber
\end{eqnarray}
\newline
where\, $\,P=MV\,$ and $\,p=mv\,$ are Momentum\,=\,Mass*Velocity of
the BP and gas atoms, respectively,\, $\,R\,$ is BP's position (and
its total path if it starts from the origin $\,R=0\,$),\, %
$\,\rho\,$ represents relative distance of atoms %
from BP,\, $\,x=\{\rho,p\}\,$,\,\, %
$\,\int \dots =\int \dots\,dx\,$,\, %
\,$\,\nu\,$ is mean gas density,\, %
$\,\Phi(\rho)\,$ is BP-atom (repulsive) %
interaction potential,

\begin{equation}
\begin{array}{l}
G_{m}({p})\,\equiv\,(2\pi Tm)^{-\,3/2}
\exp{(-{p}^{2}/2Tm)}\,\,\,,\\
E({r})\,\equiv\,\exp{[-\,\Phi({r})/T\,]}\,\,\,,\\
E^{\,\prime}( r)\equiv dE({r})/dr\, =\,-\,%
\Phi^{\,\prime}(r)\, E(r)/T\,\,\,,
\end{array}   \nonumber
\end{equation}
and, at last, $\,\psi(x)\,$ is arbitrary %
bounded function, and $\,f\,$ is the external %
force. In fact, presence of this force and %
corresponding operator %
$\,f\cdot\partial/\partial  P\,$\, %
(entering any equation of the BBGKY hierarchy)  %
is the only difference of Eq.\ref{fev} from %
equations of \cite{pro}.

Notice \cite{hs} that the operator
$\,\widehat{\mathcal{L}}\,$ can be
written in a more compact and visual form,

\begin{eqnarray}
\widehat{\mathcal{L}}\,=\,%
\int [\,1+\psi(x)\,]\,\,\widehat{L}%
\left(V,\frac {\partial}{\partial P}\right)\,%
\left[\frac {\delta }{\delta \psi(x)}%
\,+\,\nu\, g(x)\,\right]\,%
\label{fop}
\end{eqnarray}
with\,\,\,\, \,\,\,\,\, \,\,\,\,\,\,\,
\[
g(x)\,\equiv\,\,G_m(p)\,E(\rho)\, \,\,
\]
and
\begin{eqnarray}
\widehat{L}\left(V,\frac {\partial}{\partial P}\right)\,=\,%
({V}-{v})\cdot\frac {\partial}{\partial \rho}\,+\,\Phi^{\prime}(\rho)
\cdot\left(\frac {\partial}{\partial
{p}}-\frac {\partial}{\partial {P}}\right)\,%
\label{op}
\end{eqnarray}
Notice also that
\begin{eqnarray}
\widehat{\mathcal{L}}\,G_M(P)\,=\, 0\,\,\,,\label{est}
\end{eqnarray}
hence, $\,G_M(P)\,$ is equilibrium stationary %
solution to Eq.\ref{fev} at $\,f=0\,$.

As before  in \cite{pro}, let us assume that %
at initial time moment $\,t=0\,$ the BP was disposed at definitely
known point $\,R=0\,$ while the gas was in equilibrium (that is the
force $\,f\,$ %
is switched on just at $\,t=0\,$). Then initial %
condition to Eq.\ref{fev} again is
\begin{equation}
\mathcal{V}\{t=0\}\,=\,G_M({P})\,\delta(R)\,\,,\label{ic}
\end{equation}
Solving Eq.\ref{fev} with this condition, we %
can find the probability density distribution %
of BP's variables, %
$\,W=W(t,R,P\,|\nu,f)\,$, - which eventually is %
most interesting for us, - from
\[
W\,=\,\mathcal{V}\{\psi=0\}
\]
At $\,\psi\neq 0\,$, the functional %
$\,\mathcal{V}=\mathcal{V}\{t,R,P, %
\,\psi\,|\,\nu,f\}\,$ represents %
full statistics of BP's motion in a ``cloud'' %
of its correlations with surrounding %
gas atoms.

Importantly, as it was underlined in %
\cite{pppro,ppro,tmf}, the virial relations %
\cite{pro,pppro,ppro,tmf,ig,hs} remain valid %
in presence of the external force. %
Thus at any $\,t,\,R,\,P\,$ and $\,f\,$
we can write

\begin{eqnarray}
\mathcal{V}\{\sigma +\psi\,|\,\nu\}\,=\,%
\mathcal{V}\{\psi/(1+\sigma)\,|\,(1+\sigma)\,\nu\} %
\,\label{vr}
\end{eqnarray}
\newline with $\,\sigma =\,$\,const\, being %
arbitrary constant from interval %
$\,-1<\sigma <\infty\,$. %
One can easy derive this ``generating
virial relation'' directly from Eq.\ref{fev} %
and condition (\ref{ic}) by exploiting %
clearness of the form (\ref{fop}). This relation, %
(\ref{vr}),  will play a crucial role below.

For further it is convenient to go from %
$\,\mathcal{V}\,$ to corresponding characteristic %
functional $\,\overline{\mathcal{V}}\,$ %
defined by the Fourier transform
\[
\overline{\mathcal{V}}\,=\,%
\int \int \exp{(i\,k\cdot R + i\,\xi\cdot V\,)}\, %
\,\mathcal{V}\, dP\,dR \,\,\,,
\]
and to the BP's characteristic function
\[
\begin{array}{l}
\overline{W}\,=\,%
\int \int \exp{(i\,k\cdot R + i\,\xi\cdot V\,)}\, %
\,W\, dP\,dR \,=\,\\%
=\, \overline{\mathcal{V}}\{\psi=0\}\,=\,\\ %
=\, \langle\,  \exp{[i\,k\cdot R(t) + %
i\,\xi\cdot V(t)\,]} \,\,|\,\nu,f\, \rangle\,\,\,,
\end{array}
\]
where we introduced angle brackets %
$\,\langle\dots |\nu,f\rangle\,$  to denote %
statistical averaging at given $\,\nu\,$ %
and $\,f\,$ under mentioned initial conditions. %

The angle brackets' designations help to represent %
statistical contents of $\,\mathcal{V}\,$ %
 and $\,\overline{\mathcal{V}}\,$ in very %
transparent form. Namely, in accordance with the %
$\,\mathcal{V}\,$'s definition %
 \cite{pro} \,\footnote{\,

See also \cite{pppro,ppro,tmf} and, for details, %
the basic Bogolyubov's  definition \cite{bog}.}\,,

 \begin{eqnarray}
 \overline{\mathcal{V}}\,=\frac%
{\langle\,  \exp{\left[ik\cdot R(t) + %
i\xi\cdot V(t)+ \int \ln{(1+\psi(x))}\,%
 \upsilon (t,x)\,\right]} \,|\,\nu,f \rangle}%
{\langle\,  \exp{\left[ \int \ln{(1+\psi(x))}\,%
 \upsilon (t,x)\,\right]} \,\,|\,\nu,0 \rangle}%
\,\,\, \label{V}
\end{eqnarray}
Here
\[
 \upsilon (t,x)\,=\sum_j \,\,%
\delta(\rho-(r_j(t)-R(t))) \, \delta(p-p_j(t))\,
\]
is microscopic gas density in the \,$\,\mu\,$-space, and
\[
\langle\,  \exp{\left[ \int \ln{(1+\psi(x))}\,%
 \upsilon (t,x)\,\right]} \,\,|\,\nu,0\, \rangle %
\,=\,\exp{\left[\,\nu\int g(x)\,\psi(x)\,\right]}\,\,\,,
\]
thus expressing the Poissonian statistics of ideal gas in
equilibrium.

The Eq.\ref{fev} and the initial condition to it %
transform into

\begin{eqnarray}
\frac {\partial\overline{\mathcal{V}}}{\partial t} = %
k\cdot \frac {\partial\overline{\mathcal{V}}}{\partial \xi} + %
f\cdot\frac {i\xi}M \,\overline{\mathcal{V}}\,+\,%
\widehat{\mathcal{L}}^{\,\prime}\, \overline{\mathcal{V}}\, %
\,\,,%
\label{fev1}\\ %
\widehat{\mathcal{L}}^{\,\prime}\,\equiv\, %
\int [\,1+\psi(x)\,]\,\,\widehat{L}%
\left(\frac {\partial}{\partial i\xi}\,\,, %
-\frac {i\xi}M\right)\,%
\left[\frac {\delta }{\delta \psi(x)}%
\,+\,\nu\,g(x)\,\right]\,\,\,,%
\label{fop1}\\
\overline{\mathcal{V}}\{t=0\}\,=\, %
\exp{[\,-T\xi^2/2M\,]}\,\,\, \label{ic1}
\end{eqnarray}
The problem is an  adequate analysis of this equation %
from viewpoint of the characteristic function %
$\,\overline{W}\,$.

\section{Cumulant representation}

Next, it is reasonable to introduce cumulants and %
their generating functional:

\begin{eqnarray}
\overline{\mathcal{V}}\,=\, \exp{\,
\mathcal{S}}\,\,\,,\nonumber\\
\mathcal{S}\,=\, \sum_{a,b\, =\,0}^\infty \, %
\frac {(ik)^a (i\xi)^b}{a!\,b!}%
\, \mathcal{S}_{a\,b}\{t,\psi\,|\,\nu,f\}\,\,\nonumber
\end{eqnarray}
Evidently,

(i)\, any of the functionals %
$\,\mathcal{S}_{a\,b}\{t,\psi\,|\,\nu,f\}\,$, at $\,a+b>0\,$, %
represents all mutual irreducible correlations of %
$\,(a+b+n)\,$-th order between $\,a\,$ samples of %
BP's path $\,R(t)\,$, $\,b\,$ samples of BP's velocity %
$\,V(t)\,$ and $\,n\,$ samples, - with $\,n=1,2,\dots\,$, %
- of instant gas state %
$\, \upsilon (t,x)\,$. At $\,n=0\,$, %
$\,\mathcal{S}_{a\,b}\{t,\,0\,|\nu,f\}\,$ is $\,(a+b)\,$-th order %
mutual cumulant of $\,R(t)\,$ and $\,V(t)\,$ irrespective to
the instant gas state. %
Taking into account that under our initial condition %
\,$\,R(t)=\int_0^t V(t^{\,\prime})\,dt^{\,\prime} \,$,\, %
and denoting purely irreducible correlations %
(i.e. cumulants) by %
double angle brackets, we can write
\begin{eqnarray}
\mathcal{S}_{a\,b}\{t,\,0\,|\nu,f\}\,=\,\langle\langle R^{\,a}(t)\,V^b(t)\,%
|\,\nu,f\rangle\rangle\,=\, \label{cum}\\
=\, \int_0^t \dots \int_0^t %
\langle\langle\, V(t_1) \dots V(t_a)\,V^b(t)\,%
|\,\nu,f\,\rangle\rangle\, dt_1\dots dt_a\,\,\,;\nonumber
\end{eqnarray}

(ii)\, because of the relation (\ref{vr}) any of the functionals %
$\,\mathcal{S}_{a\,b}\{t,\,\psi\,|\nu,f\}\,$ possesses similar property:
\begin{eqnarray}
\mathcal{S}_{a\,b}\{t,\,\sigma+\psi\,|\,\nu,\,f\}\,=\, %
\mathcal{S}_{a\,b}\{t,\,\psi/(1+\sigma)\,|\,(1+\sigma)\,%
\nu,\,f\} \,\,\,\label{vr1}
\end{eqnarray}
with arbitrary $\,-1<\sigma=\,$const$\,<\infty\,$;

(iii)\, in terms of the functional %
$\,\mathcal{S}=\mathcal{S}\{t,k,\xi,\,\psi\,|\,\nu,f\}\,$ %
Eq.\ref{fev1} takes the form of nonlinear equation with quadratic
nonlinearity,

\begin{eqnarray}
\frac {\partial\mathcal{S}}{\partial t} = %
k\cdot \frac {\partial\mathcal{S}}{\partial \xi} + %
f\cdot\frac {i\xi}M \,+\,
\label{fev2}\\ %
+\, \int [\,1+\psi(x)\,]\,\,\widehat{L}%
\left(\frac {\partial \mathcal{S}}{\partial i\xi}  %
\,+\,\frac {\partial}{\partial i\xi}\,, -\frac {i\xi}M\right)\,%
\left[\frac {\delta \mathcal{S}}{\delta \psi(x)}%
\,+\,\nu\,g(x)\,\right]\,\,\,,%
\nonumber\\
\mathcal{S}\{t=0\}\,=\,-T\xi^2/2M\,\,\, \label{ic2}
\end{eqnarray}

Of course, eventually we are most interested in %
the generating function %
of BP's cumulants themselves, %
$\,\mathcal{S}\{t,k,\xi,\,0\,|\,\nu,f\}\,$, which %
gives diffusivity, mobility and other %
statistical characteristics of BP's motion. %
But anyway we should start from %
realizing some principal properties of the whole set of %
cumulants contained in %
$\,\mathcal{S}\{t,k,\xi,\,\psi\,|\,\nu,f\}\,$.

\section{Time evolution and spatial extension of %
BP-gas cross-correlations}

Although probability distribution %
of the BP's path $\,R(t)\,$  %
constantly evolves with time, distributions of BP's %
velocity $\,V(t)\,$ and gas state  $\,\upsilon(t,x)\,$  %
can be expected (at least, %
under proper interaction potential) to tend to a stationary %
limit.  Then
\begin{eqnarray}
\overline{\mathcal{V}}\{t,\,k=0\,,\xi,\,\psi\,|\,\nu,f\}\, %
\rightarrow\, %
\overline{\mathcal{V}}_{stat}\{\xi,\,\psi\,|\,\nu,f\}\,=\, %
\exp\,\, \mathcal{S}_{stat}\{\xi,\,\psi\,|\,\nu,f\} %
\,\,\,, \nonumber 
\end{eqnarray}
where the limit function obeys Eq.\ref{fev1} at $ \,k=0\,$ %
and zero time derivative:
\begin{eqnarray}
0\,=\, %
f\cdot\frac {i\xi}M \,\overline{\mathcal{V}}_{stat}\,+\,%
\widehat{\mathcal{L}}^{\,\prime}\,\, \overline{\mathcal{V}}_{stat}\, %
\,\, \label{st}
\end{eqnarray}
Let us discuss this equation.

The structure of the operator %
$\,\widehat{\mathcal{L}}^{\,\prime}\,$, as well as %
of $\,\widehat{\mathcal{L}}\,$, %
obviously allows solution of %
Eq.\ref{st} be extended %
from any ``good'' $\,\psi(x)\,$ (e.g. vanishing %
at $\,x\rightarrow\infty\,$)  %
to $\,\psi(x)+\sigma\,$, - with $\,\sigma\,$ being a %
constant ($\,-1<\sigma<\infty\,$), - by equating %
\begin{eqnarray}
\mathcal{S}_{stat}\{\xi,\sigma +\,\psi\,|\,\nu,f\}\,=\, %
 \mathcal{S}_{stat}\{\xi, %
\,\psi/(1+\sigma)\,|\,(1+\sigma)\nu,f\} %
\,\,\, \label{vrs}
\end{eqnarray}
Combining this virial  relation with its analogue %
(what directly following from (\ref{vr}) or (\ref{vr1})) %
for any finite $\,t\,$, we have %
\begin{eqnarray}
\mathcal{S}_{stat}\{\xi,\sigma +\,\psi\,|\,\nu,f\} \,-\, %
\mathcal{S}\{t,0,\xi,\sigma +\,\psi\,|\,\nu,f\}\, %
=\, \label{cont}\\
=\,\mathcal{S}_{stat}\{\xi, %
\,\psi^{\,\prime}\,|\,\nu^{\,\prime},f\} \,-\, %
\mathcal{S}\{t,0,\xi, %
\,\psi^{\,\prime}\,|\,\nu^{\,\prime},f\} %
\,\,\rightarrow\,0 \,\,\,, \nonumber
\end{eqnarray}
where
\[
\psi^{\,\prime}(x)\,\equiv\,\frac {\psi(x)}{1+\sigma}\,\,\,, %
\,\,\,\,\, \nu^{\,\prime}\,\equiv\, (1+\sigma)\,\nu\,\,\,
\]
The equality (\ref{cont}), in spite of its seeming triviality, %
says, together with (\ref{vrs}), %
about several important things as follow.

\,\,\,

{\it Statement 1\,}.

All stationary cross-correlation cumulamts
\[
S^{stat}_{b\,n}(x_1\dots x_n\,|\,\nu,f)\, %
\equiv\, \langle \langle \, %
V^b\,\upsilon(x_1)\dots \upsilon(x_n)\, %
|\,\nu,f\,\rangle \rangle %
\]
are different from zero and (absolutely) integrable %
functions of $\,x_j\,$. Thus all they vanish when %
any of $\,p_j\,$ or $\,\rho_j\,$ goes to infinity. %
In other words, all the BP-gas correlations are %
localized in the $\,\rho\,$-space near BP, and %
corresponding characteristic  ``volume of correlation'' %
\cite{pppro,ppro,tmf} is finite.

In essence, this property of the stationary cumulants, %
- as well as analogous property \cite{pro,pppro,ppro,tmf} of the %
arbitrary non-stationary ones,
\[
S_{a\,b\,n}(t,x_1\dots x_n\,|\,\nu,f)\, %
\equiv\, \langle \langle \, %
R^{\,a}(t)\, V^b(t)\,\upsilon(t,x_1)\dots %
\upsilon(t,x_n)\, %
|\,\nu,f\,\rangle \rangle\,\,\,, %
\]
- is necessary boundary condition for a correct %
construction of solutions to Eq.\ref{st} %
and the evolution equations.

\,\,\,

{\it Statement 2\,}.

Choosing $\,\psi(x)\,$ in Eq.\ref{cont} to be %
a ``good'' function, well localized in %
the $\,\rho\,$-space, we can conclude that  %
local values of non-stationary cumulants %
and their integrals over some of $\,x_j\,$ %
tend to their stationary limits with %
one and the same speed.

In other terms, %
none of the cumulant functions %
$\,S_{\,0\,b\,n}(t,x_1\dots x_n\,|\nu,f)\,$
can contain such a component, %
$\,c(t,x_1\dots x_n\,|\nu,f)\,$, %
that
\begin{eqnarray}
\frac {c(t,x_1\dots x_n\,|\nu^{\,\prime},f)} %
{\int c(t,x_1\dots x_n\,|\nu,f) \,dx_j} %
\,\rightarrow\, 0 %
\,\,\, \label{rest}
\end{eqnarray}
In particular, for example, %
$\,S_{\,0\,b\,1}(t,x|\nu,f)\,$ can not have %
a part, $\,c(t,x)\,$, behaving somehow  like
\begin{eqnarray}
c(t,x)\,\propto\, \frac %
{g(x)\,\theta(ut-|\rho|)}{\rho^2\,t}\, %
\rightarrow\,0\,\,\,, \,\,\,\,\,\, %
\int c(t,x)\, \rightarrow\, %
\texttt{const}\,\neq\,0\,\,\,, %
\label{rest1}
\end{eqnarray}
with $\,\theta(\cdot)\,$ being the Heaviside %
function and $\,u\,$ some characteristic velocity. %
Such (contributions to) correlations %
can be named ``phantom correlations'', for they %
simultaneously ``thaw'' at infinity and stay %
significant, leaving nonphysical %
``invisibly small BP's correlations %
with infinitely far points of gas''.

Hence,  Eq.\ref{cont} (and its parent, Eq\,\ref{vr}) %
rejects ``phantom correlations'', %
thus producing even more strong %
restrictions on possible evolution of the %
correlations than their spatial integrability, %
and ensuring finiteness of the %
``volume of correlation'' during %
all the evolution.

\,\,\,

{\it Statement 3\,}.

The two above statements naturally must be %
extended from %
$\,S_{\,0\,b\,n}(t,x_1\dots x_n\,|\nu,f)\,$ %
to general cumulants %
$\,S_{a\,b\,n}(t,x_1\dots x_n\,|\nu,f)\,$ %
including BP's path $\,R(t)\,$.

Indeed, firstly, the exclusion of the
``phantom correlations'' (possessing (\ref{rest})) %
merely determines more rigid (and physically %
meaningful) boundary conditions %
at infinity than the spatial integrability %
in itself. Secondly, %
although some of the cumulants %
$\,S_{a\,b\,n}(t,x_1\dots x_n\,|\nu,f)\,$ %
at $\,a>0\,$ definitely have no time  %
limits, all they concern the same asymptotically %
stationary state of the system as %
cumulants with $\,a=0\,$. Hence, they are %
determined by the same boundary conditions.

\,\,\,

Notice that mathematically exclusion of the %
``phantom correlations'' means commutativity %
of limits $\,t\rightarrow\infty\,$ and %
$\,\psi(x)\rightarrow\,$const\,, while %
physically finiteness of a number of gas atoma %
actually involved into BP-gas correlations. %

\,\,\,

Let us apply this to the simplest cumulants %
$\,S_{1\,0\,n}(t,x_1\dots x_n\,|\nu,f)\,$  %
and their generating functional %
$\,\mathcal{S}_{1\,0}(t,\psi\,|\nu,f)\,$ , %
and substantiate one more principal %

\,\,\,

{\it Statement 4\,}.

Correlations between the total BP's path $\,R(t)\,$ %
and instant gas state $\,\upsilon(t,x)\,$ %
generally (at $\,f\neq 0\,$) are growing with time %
under the same law as the path itself.

\,\,\,

Clearly, far enough in the stationary state, %
at $\,f\neq 0\,$, %
mean value of BP's path must grow proportionally %
to time, $\, \langle\langle R(t) \rangle\rangle = %
\langle\langle V(\infty) \rangle\rangle\,t\,+\,$const\,, %
while cross-correlation cuulants, %
$\, \langle\langle R(t)\,\upsilon(t,x_1)\dots %
\upsilon(t,x_n)\rangle\rangle\,$, %
either tend to constants or grow not faster %
than $\, \langle\langle R(t) \rangle\rangle\,$ %
(otherwise the Statement 4 is even more than true). %
Therefore, firstly, we can introduce limit
\[
\mathcal{S}_{1\,0}^{(1)}(\psi\,|\nu,f)\, \equiv\, %
\lim_{t\,\rightarrow\\,\infty}\,\, %
\frac {\mathcal{S}_{1\,0}(t,\psi\,|\nu,f)}t \,
\]
Secondly, the above Statement 3 about the boundary  %
conditions at infinity allows us to extend virial %
relations (\ref{vr1}) to the limit functional, writing %
\begin{eqnarray}
\mathcal{S}_{1\,0}^{(1)}(\sigma+\psi\,|\nu,f)\,=\, %
\mathcal{S}_{1\,0}^{(1)}(\psi/(1+\sigma)\, %
|(1+\sigma)\nu,f)\, %
\label{vrs1}
\end{eqnarray}
and
\begin{eqnarray}
\mathcal{S}_{1\,0}^{(1)}(\sigma+\psi\,|\nu,f)\,-\, %
\frac {\mathcal{S}_{1\,0}(t,\sigma+\psi\, %
|\nu,f)}t \,=\,\label{cont1}\\
=\,\mathcal{S}_{1\,0}^{(1)}(\psi^{\,\prime} %
\,|\nu^{\,\prime},f)\,-\, %
\frac {\mathcal{S}_{1\,0}(t,\psi^{\,\prime}\, %
|\nu^{\,\prime},f)}t \,\rightarrow\,0\,\,\,
\nonumber
\end{eqnarray}
(withe above defined primed variables).

The limit virial relation (\ref{vrs1}) %
unambiguously claims that the limit functional %
$\,\mathcal{S}_{1\,0}^{(1)}(\psi\,|\nu,f)\,$ %
is actually depending on $\,\psi(x)\,$, since %
otherwise (\ref{vrs1}) would claim independence %
of the stationary mean BP's drift velocity %
$\,\lim\,\, \langle\langle R(t) \rangle\rangle/t\, =\, %
\langle\langle V(\infty) \rangle\rangle\,$ %
on the gas density $\,\nu\,$ (which is obviously %
unacceptable). Hence, all the limits %
$\,\lim\,\, \langle\langle R(t)\,\upsilon(t,x_1)\dots %
\upsilon(t,x_n)\rangle\rangle/t\,$\, %
are non-zero finite quantities. This is the same as %
the Statement 4.

Simultaneously we have came to

\,\,\,

{\it Statement 5\,}.

The time-averaged drift velocity, $\,R(t)/t\,$, %
is actually random quantity regardless of duration %
of the averaging.

\,\,\,

Indeed, otherwise all the limits %
\,$\,\lim\,\, \langle\langle R(t)\,\upsilon(t,x_1)\dots %
\upsilon(t,x_n)\rangle\rangle/t\,$ %
($\,n>0\,$) would be equal to zero.

Next, we will consider fluctuations of the drift velocity.

\section{Real long-time asymptotics %
versus conventional model asymptotics, %
and BP's mobility fluctuations}

Excluding somehow from the evolution equations %
(\ref{fev}) or (\ref{fev1}) or (\ref{fev2}) the gas related field %
variable $\,\psi(x)\,$ we would obtain a closed but %
time-nonlocal equation for %
BP's variables or cumulants.. %
Such operation can be named ``exclusion of thermostat'' %
or ``derivation of a kinetic equation for BP''. %
In fact, the ``exclusion of thermostat'' %
never was realized in %
a honestly correct way. Instead, it %
always was based on {\it \,a priory} %
 neglect of inter-particle statistical %
 correlations (at least, three-particle and %
 higher-order ones). The result of this assumption %
 can be e.g. the Boltzmann-Lorentz equation \cite{re}.

Unfortunately or fortunately\,\footnote{\,

\,I think ``fortunately'' since otherwise the
world would be too primitive and boring!}\,, %
an unprejudiced investigation shows\,\footnote{\, %

See e.g. \cite{i1,tmf,p1,ig,hs1} and especially the %
Krylov's prophetical book \cite{kr}.}\, %
that in reality all the many-particle correlations %
always are significant. To feel once again what does it mean, %
we have to compare a long-time %
asymptotics of BP's cumulants
(\ref{cum}) dictated by Eq.\ref{fev2} with %
the asymptotics following from model kinetic equations, %
e.g. the Boltzmann-Lorentz equation.

Firstly, let us recall the standard

\subsection{Model asymptotics}

Kinetic equation for BP in homogeneous media looks %
like
\begin{eqnarray}
\frac {\partial W}{\partial t}\,+ \,V\cdot\frac {\partial W}{\partial
{R}}\, +\,f \cdot\frac {\partial W}{\partial {P}}\, %
=\, \widehat{\mathcal{L}}_{model}\left(V,\frac {\partial }{\partial {P}} %
\right)\, W\,\,\,, \nonumber
\end{eqnarray}
where $\,\widehat{\mathcal{L}}_{model}\,$ is kinetic operator, %
- generally integral one and certainly %
satisfying %
$\,\widehat{\mathcal{L}}_{model}\,G_M(P)=0\,$\, %
(analogue of our equality (\ref{est})), -  %
for instance, the Boltzmann-Lorentz operator (i.e. %
linearized Boltzmann operator). %
Corresponding equation for
\[
S(t,k,\xi\,|\,\nu,f)\,=\, \ln{\, %
\overline{W} (t,k,\xi\,|\,\nu,f)}\,\,\,,\nonumber
\]
- or $\,S(t,k,\xi\,|\,\nu,f)= %
\mathcal{S}(t,k,\xi, \psi=0\,|\,\nu,f)\,$ in our %
above designations, - is

\begin{eqnarray}
\frac {\partial S}{\partial t}\,= \,k\cdot %
\frac {\partial S}{\partial
{\xi}}\, +\,f \cdot\frac {i\xi}{M}\, %
+\, \widehat{\mathcal{L}}_{model}\left( %
\frac {\partial S}{\partial i\xi} +  %
\frac {\partial }{\partial i\xi}\,,-\frac {i\xi}M %
\right)\, 1\,\,\, \nonumber
\end{eqnarray}

with the same initial condition (\ref{ic2}).

Consider asymptotic behavior of the cumulants generating %
function, $\,S(t,k,\xi\,|\,\nu,f)\,$, at %
$\,t/\tau_0 \rightarrow\infty\,$, where $\,\tau_0\,$ is %
characteristic momentum relaxation time . The asymptotic %
has the well known and clear form,

\begin{eqnarray}
S(t,k,\xi\,|\,\nu,f)\, =\, %
S^{(1)}(k\,|\,\nu,f)\,t\, +\, %
S^{(0)}(k,\xi\,|\,\nu,f)\,+\,\dots\,\,\,, \label{as0}
\end{eqnarray}

where the dots replace remaining terms what are decaying %
to zero, and
\begin{eqnarray}
S^{(1)}\,= \,k\cdot %
\frac {\partial S^{(0)}}{\partial
{\xi}}\, +\,f \cdot\frac {i\xi}{M}\, %
+\, \widehat{\mathcal{L}}_{model}\left( %
\frac {\partial S^{(0)}}{\partial i\xi} +  %
\frac {\partial }{\partial i\xi}\,,-\frac {i\xi}M %
\right)\, 1\,\,\, \nonumber
\end{eqnarray}
This linear asymptotics means, obviously, that %
second- and higher-order irreducible self-correlations of BP's %
 velocity are fast decaying functions %
(decaying exponentially or at least in an %
 integrable fashion) %
 of time differences, %
 so that all cumulants (\ref{cum}) with $\,b>0\,$ %
 tend to finite constants, while at $\,b=0\,$ %
 correspondingly grow proportionally to time.

 In other words,  - borrowed from the %
 probability theory \cite{fel,luc}, - %
$\,R(t)\,$ asymptotically behaves as a %
random process with %
independent increments and thus has asymptotically %
Gaussian probability distribution.

Now, let us return to our exact equations and %
perceive that

\subsection{Exact equations forbid the %
model asymptotics}

In ideal gas, regardless of its density, certainly %
there are no collective excitations and thus no %
hydrodynamical correlations. Therefore, for the first look, %
we can expect that the Boltzmann-Lorentz equation %
gives qualitatively correct description of BP's %
motion, moreover, even quantitatively correct description %
of its asymptotical statistical properties. In other %
words, we would like to expect that %
at $\,t/\tau_0\rightarrow\infty\,$, %
similar to (\ref{as0}),

\begin{eqnarray}
\mathcal{S}\, =\, \mathcal{S}^{(1)}\,t\, +\, %
\mathcal{S}^{(0)}\,+\,\dots\,\,\, \label{as}
\end{eqnarray}

At that, $\,\mathcal{S}^{(1)}\,$ can be more formally %
introduced by $\,\mathcal{S}^{(1)}=\lim\, \mathcal{S}/t\,$. %
Again, such the asymptotics would mean that all %
irreducible self-correlations of BP's velocity, as well %
as its mutual correlations with gas state, %
$\, \upsilon (t,x)\,$, are fast enough decaying %
(integrable) functions of time differences.

This expectation, however, immediately meets serious %
objections and contradictions. We will consider them %
in a few steps.

Notice that the asymptotics (\ref{as}) %
can not be literally analogous to (\ref{as0}) since %
factor $\,\mathcal{S}^{(1)}\,$ must be essentially %
dependent on $\,\psi\,$. Indeed, from the exact %
relation (\ref{vr}) or (\ref{vr1}) it follows, %
- in view %
of arbitrariness of $\,t\,$ and %
in view of results of the previou Section 4, - that %
$\,\mathcal{S}^{(1)}\,$ must obey similar relation,

\begin{eqnarray}
\mathcal{S}^{(1)}\{k,\xi,\,\sigma +\psi\,|\,\nu,\,f\}\,=\,%
\mathcal{S}^{(1)}\{k,\xi,\,\psi/(1+\sigma)\,|\, %
(1+\sigma)\,\nu\,,f\} %
\,\,\,,\label{vr2}
\end{eqnarray}
with all potential arguments being written out. %
This relation allows $\,\mathcal{S}^{(1)}\,$ be %
independent on $\,\xi\,$, like $\,S^{(1)}\,$ in %
(\ref{as0}), but requires its actual dependence on %
$\,\psi\,$. Otherwise (excluding $\,\psi\,$ from %
the list %
of its arguments), we inevitably would come %
to conclusion %
that $\,\mathcal{S}^{(1)}\,$ is also completely %
independent on $\,\nu\,$. %
This, in turn, would imply that mean value, %
variance and higher cumulants of $\,R(t)\,$, %
and hence  BP's diffusivity and mobility, %
are completely independent on %
gas density !

Since the latter is certainly wrong statement %
(i.e. such $\,\mathcal{S}\,$ can not bring %
solution to Eq.\ref{fev2}), %
we have to accept expression %
(\ref{as}) in the form

\begin{eqnarray}
\mathcal{S}\, =\, \mathcal{S}^{(1)}\{k,\,\psi\, %
|\,\nu,f\}\,t\, +\, %
\mathcal{S}^{(0)}\{k,\xi,\,\psi\,|\,\nu,f\} %
\,+\,\dots\,\,\, \label{as1}
\end{eqnarray}
with $\,\mathcal{S}^{(1)}\,$ being a non-trivial %
functional of $\,\psi(x)\,$, so that
\begin{eqnarray}
\frac {\delta\mathcal{S}^{(1)}}{\delta\psi(x)}\, %
\neq\,0\, \,\,,\, \,\,\,\,\,\, %
\frac {\delta^2\mathcal{S}^{(1)}}{\delta\psi(x_1)\, %
\delta\psi(x_2)}\,\neq\,0\,\,\,,\, \,\,\dots\, %
\label{neq}
\end{eqnarray}
(the sequence is infinite, because %
$\, \upsilon (t,x)\,$ by its sense is %
non-Gaussian random field). Thus we came to

\,\,\,

{\it Contradiction 1\,}.

From (\ref{as1}) and (\ref{neq}) %
it follows, for large enough $\,t\,$, that
\begin{eqnarray}
\langle\langle\, R(t)\, \upsilon (t,x) \, |\,\nu,f\,\rangle\rangle %
\,\propto\,t\,\,\,, \label{rxi}\\ %
\langle\langle\, R^{\,a}(t)\, \upsilon (t,x_1) \dots %
 \upsilon (t,x_n) \, |\,\nu,f\,\rangle\rangle %
\,\propto\,t\,\,\,, \nonumber %
\end{eqnarray}
where all (omitted) coefficients on the right-hand sides %
generally (at $\,f\neq 0\,$) are essentially finite, %
that is non-zero, non-negligible and non-vanishing %
with time. In particular, in respect to the %
first row of (\ref{rxi}),
\begin{eqnarray}
t^{-1}\,\langle\langle\, R(t)\, %
 \upsilon (t,x) \, |\,\nu,f\,\rangle\rangle %
\,\rightarrow \,\,c(x|\nu,f)\,t\,\,\,, \label{c0}
\end{eqnarray}
where the (restored) coefficient\, %
$\,c(x|\nu,f)\,$\, is finite in the above sense, %
so that
\begin{eqnarray}
c(x|\nu,f)\,\neq\,0\, \label{c1}
\end{eqnarray}

Hence, %
mutual irreducible correlations (cumulants) %
of BP's velocity and gas state,
\begin{eqnarray}
\langle\langle\, V(t_1)\, \upsilon (t,x) \, |\,\nu,f\,\rangle\rangle %
\,\,\,, \nonumber\\ %
\langle\langle\, V(t_1)\dots V(t_a)\, \upsilon (t,x_1) \dots %
 \upsilon (t,x_n) \, |\,\nu,f\,\rangle\rangle %
\,\,\,, \nonumber %
\end{eqnarray}
are not fast decaying (integrable) functions of %
$\,t-t_j\,$. This in hard contradiction with what %
was assumed as a ground for (\ref{as}) and (\ref{as1}) !

Clearly, attempt to include  $\xi\,$ into list of arguments %
of $\,\mathcal{S}^{(1)}\,$ can not improve the %
situation. The matter is that just emphasized %
non-integrability of of mutual, or cross, correlations %
of velocity and gas state, t.g. %
$\,\langle\langle\, V(t_1)\, \upsilon (t,x) %
\, |\,\nu,f\,\rangle\rangle\,$, %
 necessarily implies %
non-integrability of velocity's %
self-correlations, e.g. %
$\,\langle\langle\, V(t_1)\, V(t) %
\, |\,\nu,f\,\rangle\rangle\,$, %
and, %
as the consequence, violation of the assumed %
asymptotics (\ref{as}). This will be seen soon %
in Remark 2.

\,\,\,

{\it Remark 1\,}.

Considering infinitesimal form of  relation (\ref{vr2}), %
one can easy obtain
\begin{eqnarray}
\int \langle\langle\, R(t)\, \upsilon (t,x) \, %
 |\,\nu,f\,\rangle\rangle \, %
\,=\,\nu\, \frac {\partial }{\partial \nu}\, %
\langle\langle\, R(t) \, %
 |\,\nu,f\,\rangle\rangle \,\,\,, \label{c2}
\end{eqnarray}
which is example of particular virial %
relations \cite{pro}. Combining it with (\ref{c0}), %
we have
\begin{eqnarray}
\int c(x|\nu,f)\,=\,\lim\,\,  %
\nu\, \frac {\partial }{\partial \nu}\, %
\langle\langle\, V_{drift}(t) \, %
 |\,\nu,f\,\rangle\rangle \,\,\,, \label{c3}
\end{eqnarray}
where\,\, $\,V_{drift}(t)\,\equiv\,R(t)/t\,$\,\, %
is time-averaged velocity, or ``drift velocity'',  of BP. %

\,\,\,

{\it Remark 2\,}.

Expressions (\ref{c0})-(\ref{c3}) %
together do prompt %
that the ``drift velocity'' $\,V_{drift}(t)\,$ %
is essentially random quantity even at arbitrary long %
duration of time averaging. The word ``essentially'' %
underlines that magnitude of fluctuations %
of  $\,V_{drift}(t)\,$ %
is comparable with its mean (ensemble average) %
value. %
This statement follows already from the (exact!) %
relation (\ref{c2}) %
if supplemented with reasonings expounded in %
\cite{pppro,ppro,tmf}\,\footnote{\,

A volume in the %
$\,\rho\,$-space actually contributing to left side of %
(\ref{c2})  has the order of $\,1/\nu\,$ (at least, %
when BP's mass is comparable with atom's mass).}\,. %

We can come to the same statement merely if combine %
the widely known general inequality %
(in essence, the Cauchy-Buniakowski inequality)
\[
\langle\langle A\,B\rangle\rangle ^2\,\leq %
\,\langle\langle A^2\rangle\rangle\, %
\,\langle\langle B^2\rangle\rangle\,
\]
with (\ref{c0}). %
Let $\,A=R(t)\,$ and $\,B=\int \upsilon(t,x) %
\,\chi(x)\,$,\, with %
\,$\,\chi(x)=\ln{[1+\psi(x)]}\,$\, being some suitably %
fixed function (see (\ref{V})). Then

 \begin{eqnarray}
\left[ \int \chi(x)\,\langle\langle\, R(t)\, %
 \upsilon (t,x) \,  |\,\nu,f\,\rangle\rangle \, %
\right]^2\,<\,\label{cb}\\
<\, \langle\langle\, \left[\int \chi(x)\, %
 \upsilon (t,x)\,\right]^2 \, %
|\,\nu,f\,\rangle\rangle \,
\langle\langle\, R^2(t) \, %
 |\,\nu,f\,\rangle\rangle \,\,\, \nonumber
\end{eqnarray}

Applying (\ref{c0}), we have

\begin{eqnarray}
\langle\langle\, R^2(t) \, %
|\,\nu,f\,\rangle\rangle %
\,>\, \frac {[\, %
\int c(x|\nu,f)\, \chi(x)\, ]^2} %
{\langle\langle\, \left[\int \chi(x)\, %
 \upsilon (t,x)\,\right]^2 \, %
|\,\nu,f\,\rangle\rangle }\,\, t^2 \,\,\,, \label{cb1}
\end{eqnarray}

In view of (\ref{c1}) and (\ref{c3}), evidently, %
$\,\chi(x)\,$ here always can be chosen such that the %
coefficient before $\,t^2\,$ is non-zero.
Hence, the variance of the drift velocity, %
$\, \sqrt{\langle\langle\, R^2(t) \, %
|\,\nu,f\,\rangle\rangle }\, /\,t\,$\,, %
is bounded from below by a finite quantity %
independent on the observation (averaging) time\,! %

Thus we came to

\,\,\,

{\it Contradiction 2\,}.

The inequality (\ref{cb1}) says that the second %
row of (\ref{rxi}) is in in contravention of %
the first row. Thus inequality (\ref{cb1}) %
 is principally %
incompatible with the assumed hypothetical %
asymptotics (\ref{as}) what produces both %
the rows simultaneously. %

Eventually, we arrive to

\,\,\,

{\it Conclusion\,}.

The model asymptotics (\ref{as}) is forbidden %
by the exact Eq.\ref{fev} (or Eq.\ref{fev1} or %
Eq.\ref{fev2}) as it was claimed in the title of %
the present Section. .

This means that true statistics %
of the  Brownian motion qualitatively differs from %
statistics implied by conventional model kinetic %
equations. The difference manifests that real BP %
has no certainly predictable drift velocity and %
mobility (or, in other words, its mobility possesses %
slow ``quasi-static'' fluctuations).

\section{Magnitude of the mobility fluctuations}

The same conclusion would appear if we tried to search %
for solution of Eq.\ref{fev2} in the form (\ref{as1}). %
But it would require much more tremendous %
consideration than the above one based on the virial %
relation (\ref{vr}). Nevertheless, of course, %
any quantitative calculations of statistical characteristics %
of the Brownian motion are impossible without %
direct investigation %
of the evolution equation (\ref{fev2}) (or (\ref{fev}) or %
(\ref{fev1})). In particular, calculation of the %
important cross-correlation cumulant function %
$\,c(x|\nu,f)\,$, - introduced in (\ref{c0}), - %
as a functional of the BP-atom interaction potential %
(or corresponding ``scattering matrix''). %
Since, however, methods for solving of such %
functional PDE  as %
(\ref{fev}) or (\ref{fev1}) or %
(\ref{fev2}) still are not %
developed\,\footnote{\,

Although some formal approaches were suggested in %
\cite{pro} and approximate ones in \cite{i1,p1}.}\,, %
we are forced to confine ourselves by rough estimates. %

Therefore, returning to inequality (\ref{cb1}), %
let us choose $\,\chi(x)=\phi(\rho)\,$. Then %
(\ref{cb1}) changes to

\begin{eqnarray}
\langle\langle\, R^2(t) \, %
|\,\nu,f\,\rangle\rangle %
\,>\, \frac {[\, %
\int c^{\,\prime}(\rho|\nu,f)\, \phi(\rho)\,d\rho\, ]^2} %
{\langle\langle\, \left[\int \phi(\rho)\, %
 \widetilde{\nu} (t,\rho)\,d\rho\,\right]^2 \, %
|\,\nu,f\,\rangle\rangle }\,\, t^2 \,\,\,, \label{cb2}
\end{eqnarray}
where
\[
c^{\,\prime}(\rho|\nu,f)\,\equiv\,\int %
c(x|\nu,f)\, dp\,\,\,
\]
and
\[
\widetilde{\nu} (t,\rho)\,\equiv\, %
\int \upsilon(x)\, dp\, =\, %
\sum_{\,j}\,\, \delta(\rho-(r_j(t)-R(t)))\,
\]
is random microscopic gas density in the %
configurational space. %
According to (\ref{c3}), %
\begin{eqnarray}
\int c^{\,\prime}(\rho|\nu,f)\,d\rho\,=\,\lim\,\,  %
\nu\, \frac {\partial }{\partial \nu}\, %
\langle\langle\, V_{drift}(t) \, %
 |\,\nu,f\,\rangle\rangle \,\,\, \label{c4}
\end{eqnarray}

To estimate the coefficient before $\,t^2\,$ %
in (\ref{cb2}), firstly, let us assume that %
the external force is sufficiently small, e.g. %
in the clear sense that %
$\,\zeta\equiv |f|\,V_0\tau_0/T\,\ll 1\,$\, %
with $\,V_0=\sqrt{T/M}\,$\,, while  %
the observation time is sufficiently large, %
e.g. in the sense of $\,t\gg \tau_0/\zeta^2\,$ %
(which means that drift component of the BP's %
path $\,R(t)\,$ is mach greater than its diffusive %
component\,\footnote{\,

Formulas and pictures for ``molecular %
Brownian motion'' %
under such regime were presented and discussed e.g. %
in \cite{jstat,last}.}\,). %
Then denominator in (\ref{cb2}) can be %
estimated as
\begin{eqnarray}
\langle\langle\, \left[\int \phi(\rho)\, %
 \widetilde{\nu} (t,\rho)\,d\rho\,\right]^2 \, %
|\,\nu,f\,\rangle\rangle \, \approx \, %
\nonumber\\ %
\approx\,\nu\int E(\rho)\,\phi^2(\rho)\,d\rho\, \approx %
\nu\int \phi^2(\rho)\,d\rho\,\,\,,\nonumber
\end{eqnarray}
while $\,V_{drift}(t)\,$ and %
$\,c^{\,\prime}(\rho|\nu,f)\,$ in (\ref{c4}) and %
$\,R(t)\,$ in (\ref{cb2}) represented as
\[
\begin{array}{l}
V_{drift}(t)\,=\,\widetilde{\mu}(t)\,f\,\,\,,\\
R(t)\,=\, \widetilde{\mu}(t)\,ft\,\,\,,\\
c^{\,\prime}(\rho|\nu,f)\,\equiv\, \nu %
\mu^{\,\prime}(\rho|\nu)\,f\,\,\,,
\end{array}\]
where $\,\widetilde{\mu}(t)\,\,$ plays the role of %
random ``small field''  mobility. %
After that (\ref{cb2}) reads
\begin{eqnarray}
\langle\langle\, \widetilde{\mu}^2(t) \, %
|\,\nu\,\rangle\rangle \,=\, %
\langle\, \widetilde{\mu}^2(t) \, %
|\,\nu\,\rangle - %
\langle\, \widetilde{\mu}(t) \, %
|\,\nu\,\rangle^2\,>\, \nonumber\\ %
\,>\, \frac {\nu\,[\, %
\int \mu^{\,\prime}(\rho|\nu)\, \phi(\rho)\,d\rho\, ]^2} %
{\int \phi^2(\rho)\, %
 d\rho\, } \,\,\,, \label{cb3}
\end{eqnarray}
while (\ref{c4}) turns into
\begin{eqnarray}
\int \mu^{\,\prime}(\rho|\nu)\,d\rho\,=\,  %
\frac {\partial }{\partial \nu}\, %
\langle\, \widetilde{\mu}(t) \, %
 |\,\nu\,\rangle \,\,\, \label{c5}
\end{eqnarray}
(in respect to first-order cumulants double %
and single brackets are equivalent).

Secondly, let us choose in (\ref{cb3}) such %
$\,\phi(\rho)\,$ what maximizes the right-hand %
expression. The maximization yields

\begin{eqnarray}
\langle\, \widetilde{\mu}^2(t) \, %
|\,\nu\,\rangle - %
\langle\, \widetilde{\mu}(t) \, %
|\,\nu\,\rangle^2\,>\,  %
\nu \int \mu^{\,\prime\,2}(\rho|\nu)\, d\rho\, %
\,\, \label{cb4}
\end{eqnarray}

The rest of estimate is less formal. %
Let $\,\Omega=\Omega(\nu)\,$ be a characteristic finite %
volume (space region) in the $\,\rho\,$-space, such that it %
produces main contributions to the integrals %
in (\ref{c5}) and (\ref{cb4}). Then, obviously,
\begin{eqnarray}
\int \mu^{\,\prime\,2}\, d\rho\, >\, %
\int_\Omega \,\mu^{\,\prime\,2}\, d\rho\, %
 >\, \frac {[\int_\Omega \,\mu^{\,\prime}\, d\rho\,]^2} %
 {\Omega}\,\sim\, %
\frac {[\int \mu^{\,\prime}\, d\rho\,]^2} %
 {\Omega}\,\label{es0}
\end{eqnarray}
Besides, notice that from physical point of view the only %
natural measure for the characteristic volume is the %
specific volume $\,1/\nu\,$. Therefore we have rights %
to write $\,\Omega(\nu)\sim 1/\nu\,$. %
Adding these reasonings to (\ref{c5}) and (\ref{cb4}), %
we find
\begin{eqnarray}
\langle\, \widetilde{\mu}^2(t) \, %
|\,\nu\,\rangle - %
\langle\, \widetilde{\mu}(t) \, %
|\,\nu\,\rangle^2\,\gtrsim\,  %
\left[\,\nu \, \frac {\partial }{\partial \nu}\, %
\langle\, \widetilde{\mu}(t) \, %
 |\,\nu\,\rangle\,\right]^2   %
\,\, \label{cb5}
\end{eqnarray}

A rigorous version of this estimate can be %
formulated as follows. %
For any $\,0<\alpha <1\,$ %
let $\,\Omega(\alpha,\nu)\,$ be %
the minimum of all volumes (space regions) %
satisfying
\[
\left| \int_\Omega \,\mu^{\,\prime}\, d\rho\,- %
\int \mu^{\,\prime}\, d\rho\,\right|\,<\, %
\alpha\,\left| \int\mu^{\,\prime}\, d\rho\,\right|
\]
Then one can verify that
\[
\left| \int_{\Omega(\alpha,\nu)} \,\mu^{\,\prime}\, %
d\rho\,\right|\,>\,(1-\alpha)\, %
\left| \int\mu^{\,\prime}\, d\rho\,\right|
\]
Applying this inequality in place if the last %
step in (\ref{es0}), we have
\begin{eqnarray}
\int \mu^{\,\prime\,2}\, d\rho\, >\, %
\frac {(1-\alpha)^2\,[\int \mu^{\,\prime}\, d\rho\,]^2} %
 {\Omega(\alpha,\nu)}\,\label{es}
\end{eqnarray}
Finally this yields, instead of (\ref{cb5}),
\begin{eqnarray}
\langle\, \widetilde{\mu}^2(t) \, %
|\,\nu\,\rangle - %
\langle\, \widetilde{\mu}(t) \, %
|\,\nu\,\rangle^2\,>\,  %
\left[\,\nu \, \frac {\partial }{\partial \nu}\, %
\langle\, \widetilde{\mu}(t) \, %
 |\,\nu\,\rangle\,\right]^2\,\,   %
 \max_{\alpha}\, \frac {(1-\alpha)^2}{\nu\,\Omega(\alpha,\nu)}\, %
\,\, \label{cb6}
\end{eqnarray}
This correction, however, does not cancel the %
estimate (\ref{cb5})\,\footnote{\,

About heuristic physical reasons for the estimate %
$\,\Omega(\nu)\sim 1/\nu\,$ see %
\cite{pppro,ppro,tmf}.}\,.

We see that magnitude of the mobility fluctuations %
generally can be expected on order of its mean value.

\,\,\,

{\it Remark 3\,}.

Here, to keep logics, we should %
answer two questions as follow.

Just made estimates exploited the finiteness %
(expressed by (\ref{c1})) of the function %
$\,c(x|\nu,f)\,$ which %
in turn had appeared  as a part of the %
hypothetical asymptotics (\ref{as1}). %
But the latter was logically rejected! %
Then why we can use one of its consequences? %
And why one can not imagine this function %
to be tending to zero with time  %
while its support, $\ \Omega\,$, %
in the $\,\rho\,$-space %
growing to infinity in such way that its integral %
in (\ref{c3}) stays constant? %
Under such scenario the estimate (\ref{cb6}) %
would become insignificant.

In fact, the answers already were done in Section 4. %

Firstly, asymptotics (\ref{as1}), being wrong as the whole, %
at the same time is true in respect to the %
first-order terms in $\,ik\,$-expansion of %
the $\,\mathcal{S}\{t,k,\xi,\psi|\nu,f\}\,$\,, %
more precisely, in respect to %
$\,\mathcal{S}_{\,1\,0}\{t,\psi|\nu,f\}\,$. %
Secondly, the above imaginary scenario involves %
nonphysical ``phantom correlations'' and therefore is %
forbidden by the virial relations. All this forms %
sufficient ground for our estimates.

\section{Discussion and resume}

To conclude, let us point out most principal aspects of %
our above consideration.

\,\,\,

(i)\, The exact evolution equation (\ref{fev}) for generating %
functional of many-particle %
cumulant correlation functions of the %
system ``molecular Brownian particle (BP) %
in gas'' \cite{pro,tmf} %
produces exact ``virial relations'' \cite{pro,tmf} %
connecting various statistical characteristics of the system, %
in particular, in the above considered non-equilibrium %
steady state driven by an external force applied to the BP.

\,\,\,

(ii)\, The virial relations, in their turn, %
imply obligatory requirement to all irreducible (cumulant) %
correlations between BP's velocity and total path, from %
one hand, and current microscopic gas state, from the %
another hand, to be always (even in the non-equilibrium %
steady state) integrable functions of distances between BP %
and gas atoms. Besides, all these cross-correlations must %
be located near BP, never running away %
to infinity\,\footnote{\,

In this respect it is interesting to notice that truncations %
of the BBGKY hierarchy leading to model kinetic equations (and %
thus losing the mobility fluctuations) simultaneously born %
non-local running away correlations which involve unbounded %
number of atoms.

Such correlations may look like the ``phantom correlation'' %
(\ref{rest1}). In correct theory, it would transform %
into something like %
$\,g(x)\exp{(-|\rho|/\lambda)}/\rho^2\tau_0\,$. %
This subject will be considered separately.}\,.

\,\,\,

(iii)\, As the consequence of the locality of BP-gas %
correlations, magnitude of BP's path and %
gas state cross-correlation and variance of the path are %
on the same order of value as the path itself, irrespective %
to the time of evolution and path observation.

This means that BP's mobility has no certainly predictable %
value but instead undergoes slow quasi-static fluctuations %
whose magnitude is comparable with its ensemble-average %
value.

Thus, the exact evolution equation, in opposite to %
various model kinetic equations, predicts existence %
of BP's mobility 1/f noise.

\,\,\,

(iv)\, The integrability of BP-gas correlations %
says that the latter envelope only a finite number %
of gas atoms (on order of unit) in BP's vicinity. %
Hence, neither BP nor gas remember a history of their %
interaction (conserved in states of %
far running away atoms). %
Consequently, the system has no possibilities to ``control %
and regulate'' a number of BP-atom collisions and thus BP's %
mobility.

This is just those reason of the mobility 1/f fluctuations %
what for the first was guessed in %
\cite{bk12,bk3,i1}. %

Notice, besides, that the inequality (\ref{cb6}) %
can be treated as ``uncertainty relation''%
between mean square of of the mobility fluctuations %
and the ``correlation volume''. %
At that, treating the latter  %
as a measure of of the system's memory about its %
past, we come to statement not once pronounced %
in my cited works:\, the shorter is system's %
memory, the greater is its 1/f noise. %

\,\,\,

(v)\, Importantly, all our above consideration can be %
easy generalized from BP in ideal gas to BP in %
arbitrary fluid, if using the results of %
\cite{ppro,tmf}\,(then, for instance,  %
generalization of the estimate (\ref{cb6}) will %
differ from (\ref{cb6}) by additional multiplier %
at its right-hand side,\, %
$\,T\,\partial \nu/\partial \mathcal{P}\,$\,, %
where $\,\mathcal{P}\,$ is gas pressure).

But, besides the logical analysis of the evolution %
equations (``Bogolyubov equations'' %
\cite{tmf}) and principal %
estimates of their solutions, %
we are interested also in potentially exact regular methods %
for analytical solving of these equations %
(in addition to approximate methods %
from \cite{i1,p1,p3,jstat} %
and formal ``boson representation'' from \cite{pro}), %
in order to calculate spectra and probability %
distributions of the mobility fluctuations. %

This may be very difficult but intriguing adventure.

\,\,\,

\,\,\,

-----------------------------------------------------

\,\,\,

\,\,\,

\end{document}